# Cutting voxel projector a new approach to construct 3D cone beam CT operator

Vojtěch Kulvait [a,*,1,2], Julian Moosmann [a,1,2], Georg Rose [b,3]

[a] Institute of Materials Physics, Helmholtz-Zentrum Hereon, Max-Planck-Straße 1, Geesthacht, 21502, Germany
[b] Institute for Medical Engineering and Research Campus STIMULATE, Otto von Guericke University, Universitätsplatz 2, Magdeburg, 39106, Germany

ABSTRACT

We introduce a novel class of projectors for 3D cone beam tomographic reconstruction. Analytical formulas are derived to compute the relationship between the volume of a voxel projected onto a detector pixel and its contribution to the line integral of attenuation recorded by that pixel. Based on these formulas, we construct a near-exact projector and backprojector, particularly suited for algebraic reconstruction techniques and hierarchical reconstruction approaches with nonuniform voxel grids. Unlike traditional projectors, which assume a uniform grid with fixed voxel sizes, our method enables local refinement of voxels, allowing for adaptive grid resolution and improved reconstruction quality in regions of interest. We have implemented this cutting voxel projector along with a relaxed, speed-optimized version and compared them to two established projectors: a ray-tracing projector based on Siddon's algorithm and a TT footprint projector. Our results demonstrate that the cutting voxel projector achieves higher accuracy than the TT projector, especially for large cone beam angles. Furthermore, the relaxed version of the cutting voxel projector offers a significant speed advantage, while maintaining comparable accuracy. In contrast, Siddon's algorithm, tuned to achieve the same accuracy, is considerably slower than the cutting voxel projector. All algorithms are implemented in a GPU optimized open-source framework for algebraic reconstruction.

## 1. Introduction

X-ray-based computed tomography (CT) has revolutionized both medicine and material science. Since its inception by H. Hounsfield in the 1970s [1], CT has become a standard diagnostic tool. The market for medical CT scanners is dominated by a narrow group of companies, including Siemens Healthineers, GE Healthcare, Canon, Philips, and Hitachi. Beyond medical applications, industrial and laboratory X-ray CT systems are extensively used in material science, particularly for defect detection and structural analysis. synchrotron-radiation sources are integral to this field due to the high brilliance of their X-ray beams, enabling dynamic studies and in-situ experiments with diverse applications in biology, industry, and material science. State-of-the-art X-ray microscopes at large synchrotron-radiation facilities achieve exceptional resolutions, down to 1 μm for samples up to 7 mm in diameter in micro-CT setups [2,3], and 50 nm for samples up to 50 μm in diameter in nano-CT setups [4].

Cone beam geometry refers to the use of not only the plane in which the X-ray source rotates and irradiates the sample but also the full 3D cone of X-rays with a given angular range. This geometry can be applied to all the aforementioned modalities, as the X-ray source is typically approximated as a point source with rays diverging from it. For certain CT setups, including synchrotron-radiation applications, the parallel beam or fan beam approximation is often sufficiently accurate and preferred due to its lower computational effort. Conversely, interventional devices such as C-arm CT and flat detector CT setups utilize cone elevation angles up to 12°. It is important to note that the Tuy condition [5] prohibits fully accurate reconstruction of regions with a nonzero cone elevation angle when using a circular source trajectory.

Advancements in iterative techniques such as ordered subsets [6], total variation (TV) regularization [7], and the use of learned priors [8] together with the increasing computational power driven by GPU computing, have significantly improved CT reconstruction quality while enabling lower radiation doses. However, these improvements come

* Corresponding author.
*E-mail addresses:* vojtech.kulvait@hereon.de (V. Kulvait), julian.moosmannt@hereon.de (J. Moosmann), georg.rose@ovgu.de (G. Rose).
1 This work was partially funded by the Hereon project *Holistic Data Analysis (HoliDAy)* under the I$^2$B Innovation, Information & Biologisation Fund, Germany.
2 This research was supported in part through the Maxwell computational resources operated at Deutsches Elektronen-Synchrotron (DESY), Hamburg, Germany.
3 This work was partially funded by the German Federal Ministry of Education and Research, Germany through the Research Campus STIMULATE (grant number 13GW0473A).

https://doi.org/10.1016/j.jocs.2025.102573
Received 26 November 2024; Received in revised form 28 February 2025; Accepted 9 March 2025
Available online 20 March 2025

with heightened computational demands. Medical CT scanners, often delivered as closed systems, are typically constrained by the hardware provided with the device, limiting their ability to fully exploit these advancements. In contrast, synchrotron-radiation applications, despite utilizing cutting-edge computational infrastructure, face challenges due to the overall problem sizes, which are typically orders of magnitude larger than those in medical applications. These challenges underscore the importance of developing new computational methods for CT projector construction and optimizing them for GPU-based architectures to address the growing demands for speed, precision, and efficiency across both medical and industrial CT applications.

Tomographic reconstruction recovers an unknown attenuation distribution $\mu$ from line integrals through the object measured during the acquisition. Radon [9] laid the mathematical foundation for tomographic reconstruction by introducing the Radon transform and its inversion, which in 2D is equivalent to the filtered backprojection (FBP) [10,11], classical CT reconstruction technique that is still used and improved [12,13]. FBP is an analytical method, which means that there is a closed-form formula that expresses $\mu$ as a function of projection data. For 3D cone-beam CT, analytical algorithms either approximate the reconstruction using FBP, e.g. the FDK algorithm [14], or apply the Radon transform inversion in higher dimensions for more precise formulas [5,15,16]. After discretization, a single backprojection step is typically sufficient for reconstruction, making these methods fast.

Algebraic, iterative, and most of the machine learning-based methods rely on a discretized forward projection operator $\mathbf{A} \in \mathbb{R}^{m\times n}$, which describes the relationship between voxel attenuation values in the imaged sample $\mathbf{v} \in \mathbb{R}^{n}$ and pixel values on the detector $\mathbf{p} \in \mathbb{R}^{m}$ such that

$$\mathbf{A}\mathbf{v} = \mathbf{p}. \tag{1}$$

In these methods, the goal is to compute the attenuation vector $\mathbf{v}_{min}$, which minimizes a certain functional, such as the squared residual error in the simplest case [17]. Advanced scenarios include nonlinear regularization, artifact correction, dynamic tomography, perfusion imaging, and advanced priors, as demonstrated in [8,18–21]. Major open-source packages for tomographic reconstruction include the CUDA-accelerated ASTRA toolbox [22], the CUDA-accelerated TIGRE [23], and the C++-based RTK [24]. Additionally, V. Kulvait is developing the OpenCL-accelerated software KCT CBCT [25], in which the projector described in this contribution is implemented.

### 1.1. Projectors and backprojectors

The **projector** is a program that computes the action of the forward CT operator $\mathbf{A}$. Given volumetric data $\mathbf{v}_{\mathrm{IN}} \in \mathbb{R}^{n}$, it produces the corresponding projection data $\mathbf{p}_{\mathrm{OUT}} \in \mathbb{R}^{m}$ as

$$\mathbf{p}_{\mathrm{OUT}} = \mathbf{A}\mathbf{v}_{\mathrm{IN}}. \tag{2}$$

The **backprojector** is a program that computes the action of the adjoint CT operator $\mathbf{A}^{\top}$. It takes the projection data $\mathbf{p}_{\mathrm{IN}} \in \mathbb{R}^{m}$ as input and produces the corresponding volumetric data $\mathbf{v}_{\mathrm{OUT}} \in \mathbb{R}^{n}$ as

$$\mathbf{v}_{\mathrm{OUT}} = \mathbf{A}^{\top}\mathbf{p}_{\mathrm{IN}}. \tag{3}$$

The backprojector plays a crucial role in iterative reconstruction methods by incorporating projection errors back into the reconstruction domain.

In principle, storing the precise CT operator $\mathbf{A}$ in memory could enable projection and backprojection to be performed directly as matrix multiplications. However, the size of $\mathbf{A}$ makes this approach impractical. Consider the following example of a 3D CT reconstruction problem:

- **Projection size**: 512 angular views with a $512 \times 512$ detector stored as 4-byte floats require approximately 0.5 GB of storage.
- **Volume size**: A $512 \times 512 \times 512$ volumetric grid stored as 4-byte floats also requires approximately 0.5 GB of storage.
- **Operator size**: Assuming each voxel contributes, on average, 8 voxel–pixel relationships stored as 12-byte entries (voxelID, pixelID, effectSize), the total size of the operator would be approximately 6.5 TB.

As the resolution of the volume or detector increases, the storage requirement for $\mathbf{A}$ becomes prohibitively large.

To address this challenge, modern reconstruction algorithms employ matrix-free approaches, where the projector and backprojector compute the action of $\mathbf{A}$ and $\mathbf{A}^{\top}$ on the fly. This approach takes advantage of highly parallelized GPU architectures, enabling efficient computation without the need to explicitly store $\mathbf{A}$. A prominent example of matrix-free methods is Krylov subspace methods [17,26]. While these methods are computationally efficient, they introduce challenges in operations such as preconditioning, which often rely on explicit matrix factorizations. Overcoming these challenges requires specialized strategies tailored to iterative tomographic reconstruction.

### 1.2. Pixel-driven projectors

Ray-casting or ray-tracing projectors are pixel-driven methods that iterate over individual detector pixels, with foundations in computer graphics [27–29]. The central concept involves casting a ray from the X-ray source to a specific detector pixel and determining its interactions with the voxels along its path. Using path increments smaller than the voxel size, the line integral of the attenuation coefficients is constructed additively. For each increment, the length of the increment is multiplied by the attenuation value of the voxel corresponding to the path position. These contributions are summed along the ray to yield the projection value, effectively computing the product of a row of $\mathbf{A}$ with the input vector $\mathbf{v}_{\mathrm{IN}}$.

Siddon's algorithm [30] is a ray-tracing technique that accurately computes the path length of a ray through a voxel grid by utilizing the uniform spacing of the planes separating the voxels. Instead of using arbitrary increments, the algorithm tracks the intersections of the ray with the voxel grid and calculates the distance between successive intersections in all three dimensions. The length of the ray path through each voxel is then computed based on these fixed intersection distances. Improved versions of Siddon's algorithm have been developed for various scenarios, see [31,32].

The main inaccuracy arises from the fact that casting rays to different points within a pixel produces slightly different line integrals. To improve accuracy, one can cast rays towards multiple points on the pixel surface, creating a grid of $K \times K$ equally spaced points and averaging the values. In this work, we use Siddon's algorithm to compare the accuracy of other methods, where e.g. `Siddon8` uses $8 \times 8 = 64$ rays per pixel. However, as pixel sampling increases, the computational cost grows rapidly, making it impractical for iterative reconstruction. Therefore, we explore computationally efficient alternatives that achieve a balance between accuracy and performance.

### 1.3. Voxel-driven projectors

As discussed earlier, casting rays to different locations within a pixel introduces inaccuracies. In iterative reconstruction, these inaccuracies have less impact on the projector, as we typically project continuous objects with slowly varying voxel attenuations. However, backprojectors are responsible for updating error metrics, which may not be continuous. Even subtle inaccuracies in the projector can lead to significant issues during backprojection. Additionally, pixel-driven backprojectors require voxel memory synchronization when run in parallel, further slowing down the process. For these reasons, voxel-driven projectors and backprojectors were introduced.

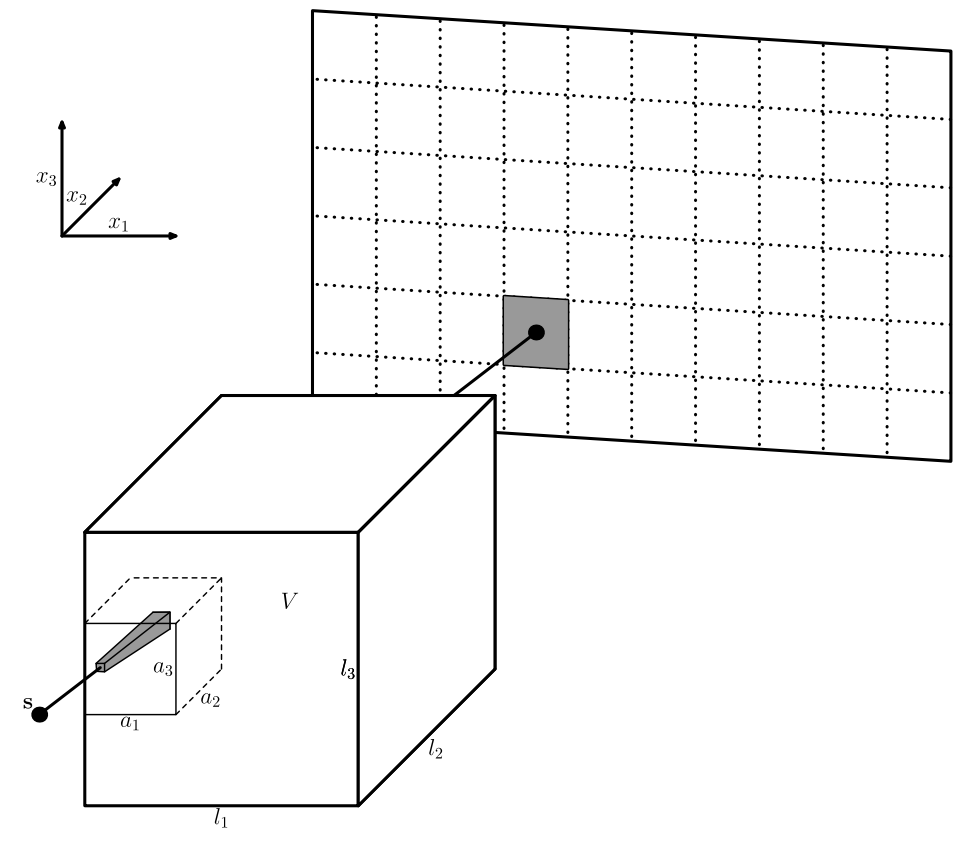

**Fig. 1.** The Cutting Voxel Projector estimates a voxel's contribution to a pixel by calculating the volume of its intersection with all rays directed toward that pixel.

Footprint projectors estimate the blur on the detector caused by each voxel, they rely on simplified geometric assumptions and are phenomenological in nature. The distance-driven (DD) projector [33] approximates this blur as a rectangular shape with constant intensity, resampled onto the detector grid. The trapezoid-rectangle (TR) projector [34] models intensity as constant in the center with linearly decaying tails, with boundaries aligned to the projections of voxel edges. The trapezoid-trapezoid (TT) projector [34] further extends this approach, approximating the footprint as the product of two trapezoidal functions in orthogonal directions of the detector grid, see also [35].

In this work, we introduce the Cutting Voxel Projector, a voxel-driven approach grounded in physical and geometrical principles. This method calculates the detector blur by determining the volume of the intersection, or "cut", formed by rays connecting a voxel and a detector pixel, as illustrated in Fig. 1. By reformulating the problem using integral calculus, we directly relate the size of the cut volume to the line integrals through the voxels. This eliminates the need to cast a large number of rays per pixel to improve accuracy and avoids reliance on simplified assumptions about footprint shapes, ensuring greater precision and flexibility.

## 2. Mathematical derivation of the cutting voxel projector

In this section, we derive a fast, nearly exact voxel-driven projector for 3D cone-beam geometry. We first define the conventions for indexing the CT volume and specifying the coordinate systems, which provide a consistent framework for the derivation. Using curvilinear calculus, we combine the line integral across the voxel cut with the surface integral over the pixel area, allowing us to compute the volume integral of the cut attenuation. This enables direct use of the cut volumes in the pixel contributions, avoiding the need for varying path lengths or approximating footprint shapes. As a result, our voxel-driven projector is based on the underlying physical model using the Lambert–Beer law, in contrast to phenomenological approaches.

For readers more focused on the implementation rather than the detailed derivation, this section can be skipped. A summary of the key formulas is provided in Section 3 on implementation.

Throughout this paper, we assume that the X-ray source, the scanned object, and the detector are located in the reference Euclidean space $\mathbb{R}^3$. For a given view, the X-ray source is positioned at $\mathbf{s} = (s_1, s_2, s_3)$, while the object remains fixed. The source and detector positions vary across views, but for clarity, we describe the parametrization relative to a single view along the tomographic trajectory.

### 2.1. Discretization of the scanning volume

The object to be scanned is contained within a rectangular box $V$, centered at the origin $(0,0,0)$ in the reference Euclidean space $\mathbb{R}^3$. The box has edge lengths $(l_1, l_2, l_3) \in \mathbb{R}^3_+$. To discretize this volume, we introduce the following parameters:

- $N_1, N_2, N_3 \in \mathbb{N}$: the number of voxels along each edge of the box,
- $a_1, a_2, a_3 \in \mathbb{R}_+$: the size of each voxel along the respective axes.

The relationship between the voxel sizes and the edge lengths is given by $a_\alpha N_\alpha = l_\alpha, \quad$ for $\alpha \in \{1,2,3\}$, so the whole scanning volume can also be described as

$$V = \{\mathbf{x} \in \mathbb{R}^3 : |x_1| < \frac{l_1}{2}, |x_2| < \frac{l_2}{2}, |x_3| < \frac{l_3}{2}\}. \tag{4}$$

Each voxel is indexed by $(i,j,k)$, where $i \in \{0,\dots,N_1-1\}$, $j \in \{0,\dots,N_2-1\}$, and $k \in \{0,\dots,N_3-1\}$. The voxel with multiindex $(0,0,0)$ is located at the corner of the box $V$ rather than its center, and its center coordinate is

$$\mathbf{x}^{(0,0,0)} = \left(-\frac{l_1}{2} + \frac{a_1}{2}, -\frac{l_2}{2} + \frac{a_2}{2}, -\frac{l_3}{2} + \frac{a_3}{2}\right). \tag{5}$$

A general voxel $(i,j,k)$ has its center at

$$\mathbf{x}^{(i,j,k)} = \mathbf{x}^{(0,0,0)} + \left(ia_1,\ ja_2,\ ka_3\right),$$

and its volume is defined as:

$$V^{(i,j,k)} = \left\{\mathbf{x} \in \mathbb{R}^3 : \left|x_\alpha - x_\alpha^{(i,j,k)}\right| < \frac{a_\alpha}{2},\ \alpha \in \{1,2,3\}\right\}. \tag{6}$$

Finally, we assume that the attenuation function $\mu : \mathbb{R}^3 \to \mathbb{R}^+_0$ is constant within each voxel. Its values, which are unknowns in the tomographic problem (1), form the vector $\mathbf{v}$ with entries $v^{(i,j,k)}$ such that

$$v^{(i,j,k)} = \mu(\mathbf{x}), \quad \forall \mathbf{x} \in V^{(i,j,k)}. \tag{7}$$

### 2.2. Discretization of the detector

The detector surface $A$ is parametrized using two Euclidean coordinates $\chi = (\chi_1, \chi_2) \in \mathbb{R}^2$. We consider a flat panel detector with rectangular pixels arranged without gaps. We use parameters $(M,N) \in \mathbb{N}^2$ to describe the number of pixels along the two detector axes and $(b_1, b_2) \in \mathbb{R}^2_+$ to define the pixel spacing along each axis. Each pixel is indexed by $(m,n)$, where $m \in \{0,\dots,M-1\}$ and $n \in \{0,\dots,N-1\}$. The center of the pixel with multiindex $(0,0)$ is located at detector origin $\chi^{(0,0)} = (0,0)$. The center of the general pixel $(m,n)$ is at

$$\chi^{(m,n)} = \chi^{(0,0)} + (mb_1, nb_2). \tag{8}$$

The surface of pixel $(m,n)$ is defined as

$$A^{(m,n)} = \{\chi \in \mathbb{R}^2 : |\chi_\alpha - \chi_\alpha^{(m,n)}| < \frac{b_\alpha}{2},\ \alpha \in \{1,2\}\}. \tag{9}$$

### 2.3. Parametrization of rays in the source-detector geometry

For a given view, where the source and detector surface are fixed in a reference Euclidean space, we need a parametrization of the rays from the source to a given position on the detector. To describe the geometry, we introduce an auxiliary Euclidean system $(\tilde{x}_1, \tilde{x}_2, \tilde{x}_3)$ centered at the source position $\mathbf{s}$. The directions of $\tilde{x}_1$ and $\tilde{x}_2$ coincide with the directions of the detector axes $\chi_1$ and $\chi_2$, while $\tilde{x}_3$ is orthogonal to the detector plane, as illustrated in Fig. 2. This system in turn induces a spherical coordinate system $(r, \theta, \varphi)$, where:

- $r$ is the distance from the source,
- $\theta \in [0, \pi/2]$ is the polar or cone angle between the ray and the direction orthogonal to the detector plane, and
- $\varphi \in [0, 2\pi)$ is the azimuthal angle, with $\varphi = 0$ aligned with the $\chi_1$-axis of the detector plane.

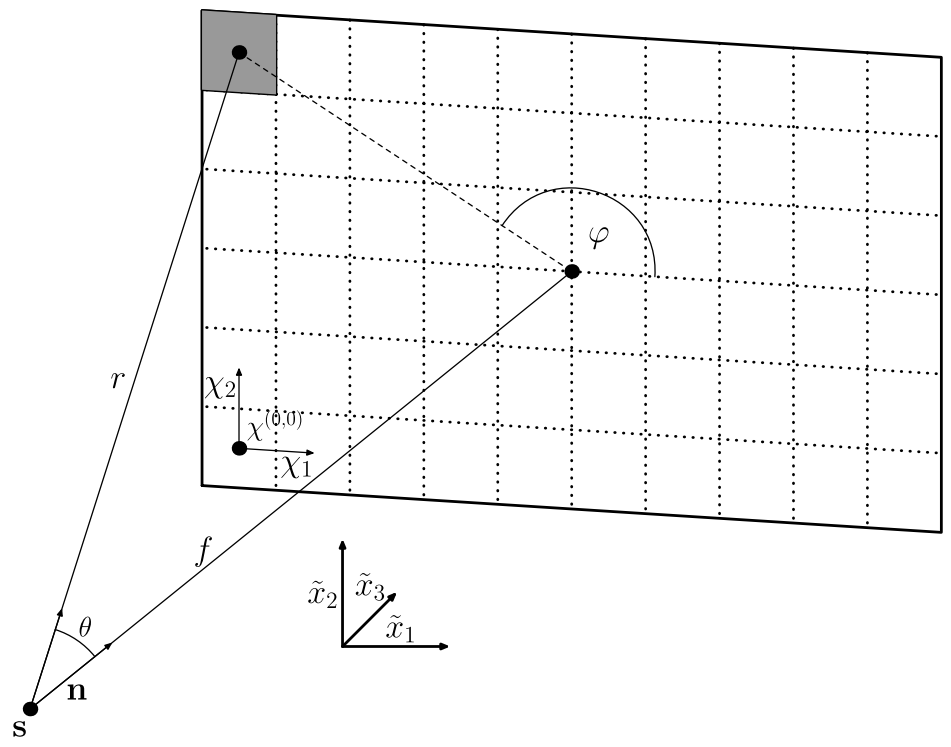

**Fig. 2.** Auxiliary local Euclidean coordinates for a given view that induce a local spherical coordinate system. Coordinate $\theta$ is the cone angle.

The principal ray, which is orthogonal to the detector plane, corresponds to $\theta = 0$. Its intersection with the detector is referred to as the principal point, denoted by $\chi^P = (\chi_1^P, \chi_2^P)$. The rays passing through the source can be parametrized in two equivalent ways:

1. By spherical coordinates $(\theta, \varphi)$ of the auxiliary system.
2. By detector coordinates $(\chi_1, \chi_2)$, which specify where the ray intersects the detector surface.

To explicitly parametrize rays in the reference Euclidean space $\mathbb{R}^3$, we proceed as follows. First, the spherical coordinates define a unit direction vector in the auxiliary Euclidean system:

$$\mathbf{d}(\theta, \varphi) = (\cos\varphi \sin\theta, \sin\varphi \sin\theta, \cos\theta).$$

This direction vector $\mathbf{d}(\theta, \varphi)$ is then rotated into the reference Euclidean system using a view-dependent orthogonal $3 \times 3$ rotation matrix $\mathbf{Q}$, which aligns the axes of the auxiliary system with those of the reference system. The resulting direction vector in the reference system is

$$\mathbf{d}_{\mathrm{ref}}(\theta, \varphi) = \mathbf{Q}\mathbf{d}(\theta, \varphi).$$

Finally, the ray in the reference system can be parametrized as

$$\mathbf{L}[\theta, \varphi](r) = \mathbf{s} + r \cdot \mathbf{d}_{\mathrm{ref}}(\theta, \varphi), \quad r \in [0, \infty), \tag{10}$$

where $\mathbf{s}$ is the source position. To parametrize the rays by detector coordinates $\chi = (\chi_1, \chi_2)$, the angular coordinates can be expressed as $\theta = \theta(\chi)$ and $\varphi = \varphi(\chi)$. Substituting these into (10) we get

$$\mathbf{L}[\chi](r) = \mathbf{s} + r \cdot \mathbf{d}_{\mathrm{ref}}(\theta(\chi), \varphi(\chi)), \quad r \in [0, \infty). \tag{11}$$

The parametrizations by detector coordinates $\chi$ and auxiliary angular coordinates $(\theta, \varphi)$ can thus be used interchangeably, depending on the context. In the following text, we utilize either form as appropriate to simplify derivations and explanations.

### 2.4. X-ray attenuation model and the X-ray transform

The Lambert–Beer law provides a physical model for the attenuation of X-rays and serves as the underlying foundation for tomographic reconstruction. Consider a ray in the direction $[\theta, \varphi]$. According to the Lambert–Beer law, the X-ray intensity measured on the detector after passing through the scanned object is given by

$$I(\theta, \varphi) = I_0(\theta, \varphi) \exp\left(-\int_0^\infty \mu(\mathbf{L}[\theta, \varphi](r))\,\mathrm{d}r\right), \tag{12}$$

where $I_0(\theta, \varphi)$ is the flat-field intensity that would be measured at the same detector position in the absence of the scanned object. The quantity

$$E(\theta, \varphi) \overset{1}{=} \ln(I_0(\theta, \varphi)) - \ln(I(\theta, \varphi)) \overset{2}{=} \int_0^\infty \mu(\mathbf{L}[\theta, \varphi](r))\,\mathrm{d}r \tag{13}$$

is referred to here as **extinction**. Note that this terminology is not standard and extinction is sometimes called the negative logarithm of transmission or the line integral of attenuation. In the context of the discretized tomographic problem (1), the extinction is often referred to as the right-hand side, the projection vector, or simply the vector $\mathbf{p}$.

### 2.5. Discretization of the X-ray transform

For a discretized volume $V$, computing the X-ray transform to provide discretized pixel-level data requires adapting the continuous relationship in (13) to the specific geometry of the CT system. This process involves determining how individual voxels within $V$ contribute to the attenuation observed at each detector pixel. In this section, we explore various methods for discretizing the X-ray transform and calculating the pixel-level projection $E^{(m,n)}$.

We introduce the length of the path of the ray passing through the voxel $V^{(i,j,k)}$ to the detector position $\chi$, given by

$$\mathcal{D}(\chi, i, j, k) = |V^{(i,j,k)} \cap \mathbf{L}[\chi](.)| = \int_0^\infty I^{i,j,k}(r, \chi)\,dr, \tag{14}$$

where $I^{i,j,k}(r, \chi)$ is a scalar indicator function defined as:

$$I^{i,j,k}(r, \chi) = \begin{cases} 1, & \text{if } \mathbf{L}[\chi](r) \cap V^{(i,j,k)} \neq \emptyset, \\ 0, & \text{otherwise.} \end{cases} \tag{15}$$

The single-ray approximation of $E^{(m,n)}$ involves casting a single ray toward the center of each pixel $\chi^{(m,n)}$ and computing the projection as

$$E_1^{(m,n)} = \sum_{i,j,k} \mu^{i,j,k} \mathcal{D}(\chi^{(m,n)}, i, j, k). \tag{16}$$

While computationally efficient, this method is prone to inaccuracies as it neglects variations across the pixel surface. In the multi-ray approximation, multiple rays are cast toward a grid of $K \times K$ points on the pixel surface. The extinction values are averaged to obtain the final result $E_{K\times K}^{(m,n)}$. While this method reduces error, its computational cost grows quadratically with $K$. In the limit with $K \to \infty$, this approach leads to the accurate integral formula

$$E_\infty^{(m,n)} = \frac{e^{(m,n)}}{a^{(m,n)}}, \text{ where} \tag{17}$$

$$e^{(m,n)} = \sum_{i,j,k} \mu^{i,j,k} \int_{\chi \in A^{(m,n)}} \mathcal{D}(\chi, i, j, k)\,\mathrm{d}S(\chi), \tag{18}$$

is the total extinction over the area of the pixel

$$a^{(m,n)} = \int_{\chi \in A^{(m,n)}} 1\,\mathrm{d}S(\chi). \tag{19}$$

Here, $\mathrm{d}S(\chi)$ represents the surface element for integration over the pixel.

From the perspective of cone beam geometry and the projective nature of the problem, mapping pixels to segments of the unit ball provides a natural way to express the limit of $E^{(m,n)}$, alternative to (17). We use the angular coordinates $(\theta, \varphi)$ from Section 2.3 as a discretization of the unit ball, centered at the source position $\mathbf{s}$. The projection in (13) is then computed over the segment of the unit ball, $(\theta, \varphi)$ corresponding to a given pixel. The formula for per pixel attenuation analogous to (17) is then

$$\bar{E}_\infty^{(m,n)} = \frac{\bar{e}^{(m,n)}}{\bar{a}^{(m,n)}}, \tag{20}$$

where

$$\bar{e}^{(m,n)} = \sum_{i,j,k} \mu^{i,j,k} \int_{(\theta,\varphi) \in A^{(m,n)}} \mathcal{D}(\theta, \varphi, i, j, k) \sin\theta\,\mathrm{d}\theta\,\mathrm{d}\varphi \tag{21}$$

is the total extinction over the area of the unit ball segment, and

$$\bar{a}^{(m,n)} = \int_{(\theta,\varphi) \in A^{(m,n)}} \sin\theta\,\mathrm{d}\theta\,\mathrm{d}\varphi. \tag{22}$$

Here, we utilize the formula for the surface element of the unit ball, which is given by $\mathrm{d}S = \sin\theta\,\mathrm{d}\theta\,\mathrm{d}\varphi$. This reformulation leverages spherical geometry to compute the contributions of each voxel to the observed attenuation at the detector pixel.

### 2.6. Cutting voxel projector

To derive the Cutting Voxel Projector X-ray transform, we define voxel cuts $V^{(i,j,k)}_{(m,n)}$ as intersections of the 3D voxel with all rays emitted from the source toward the detector pixel, see Fig. 1. The volume of the cut is

$$\mathcal{V}^{(i,j,k)}_{(m,n)} = |V^{(i,j,k)}_{(m,n)}|. \tag{23}$$

In Section 2.5, we derived the integral formulas (17) and (20) for the projector by employing a ray-casting approach and considering the limit where the number of rays per pixel approaches infinity. These surface integrals, dependent on the partial lengths $\mathcal{D}(\chi, i, j, k)$ of the rays through voxels, are rewritten here as volume integrals over voxel cuts, eliminating the explicit dependence on line integrals.

The formula (18) for projector (17) contains the pixel-voxel contribution

$$e^{(m,n)}_{i,j,k} = \int_{\chi \in A^{(m,n)}} \mathcal{D}(\chi, i, j, k)\, \mathrm{d}S(\chi). \tag{24}$$

Assuming the source-to-pixel distance $R$ and the polar angle $\theta$ are constant within each pixel, we use their values at the pixel center. The surface element in spherical coordinates $\mathrm{d}\tilde{S}(\chi) = R^2 \sin\theta\, \mathrm{d}\theta\, \mathrm{d}\varphi$ is related to the detector surface element $\mathrm{d}S(\chi)$ using $\mathrm{d}\tilde{S}(\chi) = \cos\theta\, \mathrm{d}S(\chi)$ to account for differences in surface normal angles. Substituting these formulas into (24) gives

$$e^{(m,n)}_{i,j,k} = \int_0^\infty \int_{(\theta,\varphi)\in A^{(m,n)}} \mathbf{I}^{i,j,k}(r,\theta,\varphi) R^2 \frac{\sin\theta}{\cos\theta}\, \mathrm{d}\theta\, \mathrm{d}\varphi\, \mathrm{d}r. \tag{25}$$

The volume element in spherical coordinates is:

$$\mathrm{d}V = r^2 \sin\theta\, \mathrm{d}\theta\, \mathrm{d}\varphi\, \mathrm{d}r, \tag{26}$$

so the integral (25) can be rewritten as:

$$e^{(m,n)}_{i,j,k} = \int_{V^{(i,j,k)}_{(m,n)}} \frac{R^2}{r^2 \cos\theta}\, \mathrm{d}V. \tag{27}$$

Using a source to detector distance $f$, where $R\cos\theta = f$ and assuming the source to cut distance $r$ can be taken as constant within each cut $r = r_{i,j,k,m,n}$ the formula (27) simplifies to

$$e^{(m,n)}_{i,j,k} = \frac{f^2}{r^2_{i,j,k,m,n} \cos^3\theta} \mathcal{V}^{(i,j,k)}_{(m,n)}. \tag{28}$$

This yields the discrete extinction formula for the Cutting Voxel Projector:

$$E^{(m,n)}_{CVP} = \frac{f^2}{a^{(m,n)} \cos^3\theta} \sum_{i,j,k} \frac{\mu^{i,j,k}}{r^2_{i,j,k,m,n}} \mathcal{V}^{(i,j,k)}_{(m,n)}. \tag{29}$$

For parallel ray geometry, the formula (17) simplifies to

$$E^{(m,n)}_{PR} = \frac{1}{a^{(m,n)} \cos\hat{\theta}} \sum_{i,j,k} \mu^{i,j,k} \mathcal{V}^{(i,j,k)}_{(m,n)}, \tag{30}$$

where $\hat{\theta}$ is the angle between the rays and the detector normal. For $\hat{\theta} = 0$, we have $\cos\hat{\theta} = 1$.

On the unit sphere, the integral formula (21) for the pixel voxel contribution is

$$\bar{e}^{(m,n)}_{i,j,k} = \int_{(\theta,\varphi)\in A^{(m,n)}} \mathcal{D}(\theta, \varphi, i, j, k) \sin\theta\, \mathrm{d}\theta\, \mathrm{d}\varphi, \tag{31}$$

employing (14), we obtain

$$\bar{e}^{(m,n)}_{i,j,k} = \int_0^\infty \int_{(\theta,\varphi)\in A^{(m,n)}} \mathbf{I}^{i,j,k}(r,\theta,\varphi) \sin\theta\, \mathrm{d}\theta\, \mathrm{d}\varphi\, \mathrm{d}r. \tag{32}$$

Using formula (26) for volume element, this becomes

$$\bar{e}^{(m,n)}_{i,j,k} = \int_{V^{(i,j,k)}_{(m,n)}} \frac{1}{r^2}\, \mathrm{d}V \approx \frac{1}{r^2_{i,j,k,m,n}} \mathcal{V}^{(i,j,k)}_{(m,n)}. \tag{33}$$

The area of the unit sphere projected onto a given pixel in (22) $\bar{a}^{(m,n)}$ can be computed using spherical trigonometry

$$\bar{a}^{(m,n)} = \alpha + \beta + \gamma + \delta - 2\pi, \tag{34}$$

where $\alpha, \beta, \gamma, \delta$ represent the angles at the four vertices of the spherical polygon, see [36, p. 98–99], [37]. For unit vectors $\mathbf{t}^{(m,n)}_0, \mathbf{t}^{(m,n)}_1, \mathbf{t}^{(m,n)}_2, \mathbf{t}^{(m,n)}_3$ in the direction of pixel corners (ordered counterclockwise), the area can be computed as

$$\bar{a}^{(m,n)} = 2\pi - \sum_{i=0}^{3} \arccos \frac{(\mathbf{t}_i \times \mathbf{t}_{i+1}) \cdot (\mathbf{t}_{i+1} \times \mathbf{t}_{i+2})}{|\mathbf{t}_i \times \mathbf{t}_{i+1}||\mathbf{t}_{i+1} \times \mathbf{t}_{i+2}|}. \tag{35}$$

The Eq. (17) for the Cutting Voxel Projector on unit ball is

$$\bar{E}^{(m,n)}_{CVP} = \frac{1}{\bar{a}^{(m,n)}} \sum_{i,j,k} \frac{\mu^{i,j,k}}{r^2_{i,j,k,m,n}} \mathcal{V}^{(i,j,k)}_{(m,n)}. \tag{36}$$

Formula (29) defines the cutting voxel projector for cone beam geometry, while (36) represents its version transformed to the unit ball. Similarly, (30) establishes the projector for parallel ray geometry. This concludes the theoretical derivation of the cutting voxel projector. In the next section, we describe our implementation and calculation of voxel–pixel cut volumes.

## 3. Implementation of the cutting voxel projector

The cutting voxel projector is part of a larger CT reconstruction framework written in C++ and OpenCL [25], developed by V. Kulvait. Table 1 lists the implementations of projectors and backprojectors. The software is open-source, available at https://github.com/kulvait/kct_cbct, released under the GNU GPL3 license.

For clarity and completeness, we restate here the key formulas for the different versions of the cutting voxel projector. These equations form the basis of the implementation:

- **Cone Beam Geometry** The cutting voxel projector for cone beam geometry is given by:

$$E^{(m,n)}_{CVP} = \frac{f^2}{a^{(m,n)} \cos^3\theta} \sum_{i,j,k} \frac{\mu^{i,j,k}}{r^2_{i,j,k,m,n}} \mathcal{V}^{(i,j,k)}_{(m,n)}, \tag{37}$$

  where $a^{(m,n)}$ is the area of the pixel, $f$ is source to detector distance, $\theta$ is the cone angle for the given pixel in the auxiliary geometry shown in Fig. 2, $r_{i,j,k,m,n}$ is the source-to-cut distance, and $\mathcal{V}^{(i,j,k)}_{(m,n)}$ represents the volumes of the voxel cuts.
- **Cone Beam Geometry with Transformed Detector** The cutting voxel projector for cone beam geometry, with detector transformed to the unit ball, is given by:

$$\bar{E}^{(m,n)}_{CVP} = \frac{1}{\bar{a}^{(m,n)}} \sum_{i,j,k} \frac{\mu^{i,j,k}}{r^2_{i,j,k,m,n}} \mathcal{V}^{(i,j,k)}_{(m,n)}, \tag{38}$$

  where $\bar{a}^{(m,n)}$ is the projected area of the pixel on the unit sphere given by (35).
- **Parallel Ray Geometry** For parallel ray geometry, the projector is given by:

$$E^{(m,n)}_{PR} = \frac{1}{a^{(m,n)} \cos\hat{\theta}} \sum_{i,j,k} \mu^{i,j,k} \mathcal{V}^{(i,j,k)}_{(m,n)}, \tag{39}$$

  where $\hat{\theta}$ is the angle between the rays and the detector normal.

From the formulas (37), (38), and (39), it is evident that calculating the cut volumes is central to the implementation. This chapter begins by presenting the basic algorithm used for cut volume calculations, followed by a discussion on elevation correction, which refines these computations. The implementation section then provides additional details and concludes with the pseudocode for the projector calculation.

**Table 1**
List of Implementations of CBCT and PBCT Projectors and Backprojectors, these files are in `opencl` folder of the Git repository [25]. CBCT refers to Cone Beam Computed Tomography geometry, PBCT refers to Parallel Beam Computed Tomography geometry, and PBCT2D refers to slice-wise Parallel Beam geometry. CVP stands for Cutting Voxel Projector, and `barrier` refers to local memory-accelerated and synchronized implementations.

| File Name (`.cl`) | Geometry | Description |
|---|---|---|
| `projector_cbct_cvp` | CBCT | Cutting Voxel Projector. |
| `projector_cbct_cvp_barrier` | CBCT | Cutting Voxel Projector with local memory acceleration and synchronization. |
| `backprojector_cbct_cvp` | CBCT | Cutting Voxel Backprojector. |
| `projector_cbct_siddon` | CBCT | Siddon Projector. |
| `backprojector_cbct_siddon` | CBCT | Siddon Backprojector. |
| `projector_cbct_tt` | CBCT | TT Projector. |
| `backprojector_cbct_tt` | CBCT | TT Backprojector. |
| `pbct2d_cvp` | PBCT2D | Cutting Voxel Projector/Backprojector. |
| `pbct2d_cvp_barrier` | PBCT2D | Cutting Voxel Projector/Backprojector with local memory acceleration and synchronization. |
| `pbct_cvp` | PBCT | Cutting Voxel Projector/Backprojector. |
| `pbct_cvp_barrier` | PBCT | Cutting Voxel Projector/Backprojector with local memory acceleration and synchronization. |

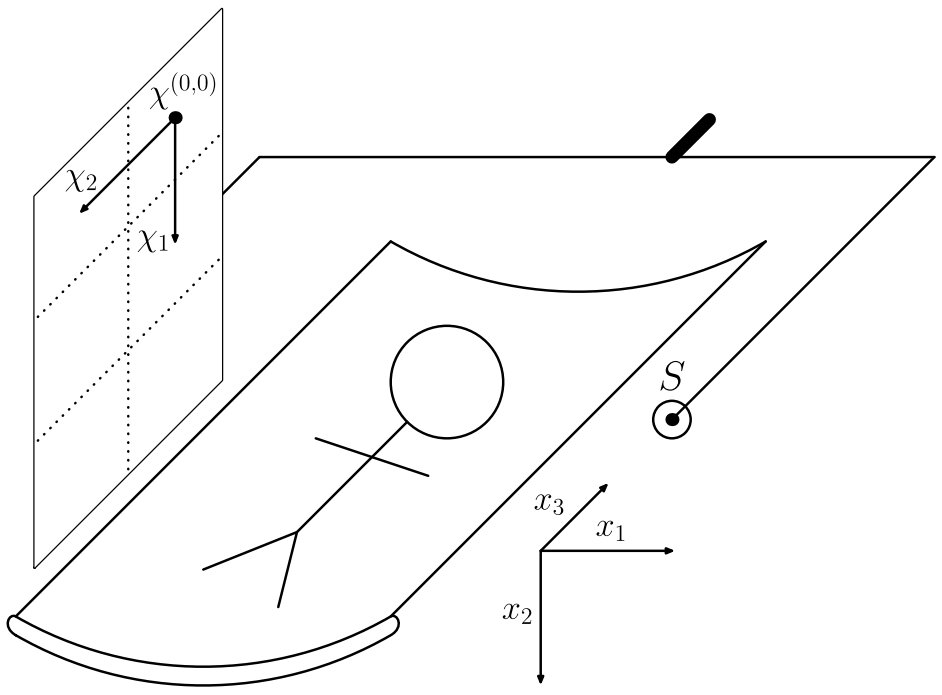

**Fig. 3.** Geometric convention and simplification for the classical cone beam CT arrangement. The axis of rotation is parallel to the $x_3$ axis of world coordinates. In addition, the $x_3$ axis is parallel to the $\chi_2$ axis of the detector, but they have opposite orientations. The origin of the detector is placed in the center of the corner pixel. This arrangement is also assumed in the software.

### 3.1. Calculating voxel cuts

To simplify the calculation of voxel cuts, current implementation assumes a geometric arrangement where the $\chi_2$ axis of the detector is parallel to the $x_3$ axis of the reference Euclidean space. By convention, their directions are opposite, as illustrated in Fig. 3. This setup is commonly used in CT trajectories, including circular and spiral trajectories, and these assumptions are fundamental to the separability of footprint projectors [34].

We consider the voxel as a 3D object, where cuts by planes collinear with the $x_3$ direction, which separate detector pixels with different $n$ indices in the $\chi_1$ direction, will have identical projections onto the voxel's $x_1x_2$ plane, regardless of the $x_3$ coordinate. This observation allows us to first calculate the cuts of the voxel base, defined by the boundaries of the pixels with increasing $n$ indices. After determining the projections onto the $x_1x_2$ plane, we extend the calculation to the third dimension by considering the projections of the cut centroids onto the pixel boundaries in the $\chi_2$ direction, corresponding to the increments of the $m$ index.

Based on this, for each voxel $V^{i,j,k}$ and each coordinate $n$ on the detector, we calculate the polygonal cuts of the voxel base, which is parallel to the $x_1x_2$ plane, intersected by rays directed to pixels in the $n$th column of the projector. Specifically, our implementation begins by determining the minimum and maximum projections, $\chi_1^{min}$ and $\chi_1^{max}$, for the voxel base. These projections are used to define the permissible range of integer values $n$ corresponding to the detector column indices onto which the voxel projects. For each $n$, we then compute two polygons: $H^{-i,j}_n$ and $H^{+i,j}_n$. The first corresponds to the intersection of the voxel base for projections in the range $\chi_1 \in [\chi_1^{min}, n-0.5]$, while the second corresponds to projections in the range $\chi_1 \in [\chi_1^{min}, n+0.5]$. The polygon $H_n^{i,j}$ is then obtained as the difference of two polygons $H_n^{i,j} = H^{+i,j}_n \setminus H^{-i,j}_n$. The area $A_n^{i,j}$ of this polygon is calculated, and the position of its center of mass is determined.

Once the area and centroid are computed, we extend the cut to the $x_3$ dimension by determining the breakpoints along the $x_3$ axis, which correspond to projections onto the pixel boundaries between the detector rows. The length of the voxel cut in the $x_3$ direction is estimated by $d^{(i,j,k)}_{(m,n)}$, which represents the distance between two consecutive breakpoints along the $x_3$ axis corresponding to the $m$th row of the detector, restricted to the admissible range of $x_3$ coordinates for the given voxel. On this basis, we estimate the volume of the voxel cut as:

$$\mathcal{V}^{(i,j,k)}_{(m,n)} = A_n^{i,j} \cdot d^{(i,j,k)}_{(m,n)}. \tag{40}$$

#### 3.1.1. Elevation correction

In the previous section, we approximated the volume of a voxel cut projected onto a pixel $(m,n)$ as the product of the voxel cut area $A_n^{i,j}$ in the $x_1x_2$ plane and the length of the line segment $d^{(i,j,k)}_{(m,n)}$ in the $x_3$-coordinate direction, which corresponds to the projection of its center of mass onto the pixel boundaries in the $\chi_2$ direction. However, in cone beam geometry, the distance $d^{(i,j,k)}_{(m,n)}$ varies across the voxel cut volume due to the divergence of rays originating from a point source. Specifically, this distance is shorter for regions closer to the source and longer for regions farther away due to diverging rays in the cone beam geometry. Even small discrepancies in this distance can lead to errors in the projector, particularly at larger elevation angles. To account for these effects, we introduce an elevation correction that leverages skew projections of voxel cuts.

To formalize the elevation correction, we introduce the elevation angle $\psi$. This angle represents the deviation between two rays. First connecting the source to a given position $(\chi_1, \chi_2)$ on the detector and the other connecting the source to the point $(\chi_1, \chi_2^P)$ in the row of the detector $\chi_2 = \chi_2^P$, which intersects principal point $\chi^P$. For a circular trajectory, this row corresponds to the intersection of the trajectory plane with the detector, and the angle $\psi$ describes the elevation of the given pixel above the trajectory plane.

In the current implementation, we apply the elevation correction only to the topmost and bottommost pixel rows associated with the voxel cut. This is because the compensatory effects of corrections for intermediate rows largely cancel each other out, as the voxel attenuation remains constant. To compute the correction, both the elevation angle $\psi$ and the shape of the voxel cut in the $x_1x_2$ plane are taken into account. This allows for a heuristic estimate of the skewed projection of the voxel cut, which balances computational efficacy and accuracy.

### 3.2. Implementation of the cutting voxel projector

This section outlines the implementation of the cutting voxel projector, which is based on the algorithm for computing voxel cuts and one of the formulas (37), (38), or (39) that describe the projector's behavior. The basic algorithm is as follows

**Algorithm 1** Cutting Voxel Projector Algorithm.

```
Concurrent Processing of Voxels:
for all voxels with coordinates (i, j, k) do
  Retrieve the attenuation value of the voxel.
  Scale the voxel contribution by 1/r^2 (for CBCT projectors).
  Determine the x1x2-plane corners of the voxel.
  Identify the corner with the minimal projection onto χ1.
  Retrieve the polygons corresponding to the pixel boundaries.
  for all polygons do
    Compute the polygon's center of mass and area.
    Calculate x3 sections corresponding to pixel χ2 boundaries.
    Get cut volume as the polygon area × length of χ2 intersection.
    Adjust the attenuation contributions based on the size of the cut.
    Optionally: Elevation correction for top and bottom pixels.
    Add the scaled contributions to the corresponding pixels.
    Use atomic operations to synchronize pixel additions.
  end for
end for
Concurrent Processing of Pixels:
for all pixels with coordinates (m, n) do
  Scale pixel values according to (37) or (38).
end for
```

The backprojector uses the same scaling logic as the projector but reverses the direction of updates. Instead of updating pixels based on voxel values, it updates voxel values based on pixel values. This design eliminates the need for atomic operations, as voxel contributions are accumulated sequentially in an aggregating variable during the loop processing of cuts. The final voxel value is written only at the end of the procedure, making the backprojector faster than the projector.

Pixel scaling factors that modify voxel sums in Eqs. (37) and (38) are independent of voxels. Thus, they can be precomputed in a separate loop through pixels. In the current implementation [25], this precomputation is employed and takes place before the backprojector loop and after the projector loop. By default is used the scaling based on unit ball intersections in the transformed geometry, as described by formula (38). It can also be explicitly activated with the `--exact-scaling` switch. Alternatively, the cosine scaling, as defined by formula (37), can be enabled using the `--cos-scaling` switch.

The initial implementation of the cutting voxel projector utilized double-precision (`float64`) arithmetic for calculating cuts and contributions. However, on NVIDIA architectures, it became clear that switching to single-precision (`float32`) arithmetic offered significant performance improvements. Additionally, enabling the `cl-fast-relaxed-math` flag in OpenCL provided further speedups with negligible precision loss, see Section 5.2. As a result, we now routinely use a projector version that combines single-precision computations with the `cl-fast-relaxed-math` setting for voxel contribution calculations. This version can be activated using the `--relaxed` switch.

Numerous further geometrical transforms were used in the current implementation, Table 1, to improve speed. In the implementation of cone beam CT geometry, the pinhole camera model and camera matrices are used to map points from world coordinates $\mathbf{x}$ to detector coordinates $\chi$ by means of projective transform. By shifting the world coordinate origin to the source point, the transform is simplified to a $3 \times 3$ matrix instead of the standard $3 \times 4$, reducing computational cost. Additionally, the detector array is stored in column-major order for better memory alignment when traversing along the $\chi_2$ axis.

### 3.3. Leveraging local memory in OpenCL

Another critical optimization is the efficient use of local memory within OpenCL, which offers faster access compared to global GPU memory. This optimization groups voxels into work-groups, allowing for quicker data access and synchronization during computation. In OpenCL, each work-item represents a thread that processes an individual voxel or pixel. A work-group consists of a set of work-items that execute on the same compute unit and share local memory. This shared memory facilitates faster data sharing and synchronization, while the locality of execution within a work-group can be enforced via barrier calls.

To further speed up the projector, we use the `--barrier` switch, enabling two-stage synchronization within the work-groups and later in global GPU memory. In the current implementation, a work-group is formed by aggregated voxels with dimensions $LN_1 \times LN_2 \times LN_3$, where each dimension typically contains $2 - 32$ voxels. These dimensions can be adjusted and optimized based on the problem and computational architecture. The contributions from these voxels are first accumulated in local memory, which is mapped to the area of the detector where the aggregated voxels are projected. Local memory ensures faster atomic operations but requires barrier synchronization. Once all contributions within a work-group are aggregated, they are written back to global memory, minimizing costly global atomic operations. This two-phase approach—local aggregation followed by global update—provides a significant speedup on most GPU architectures, where local memory has much lower latency than global GPU memory.

Applying advanced parallelization techniques to footprint projectors, particularly the TT projector, could also achieve substantial speedups. Original algorithm described in Long et al. [34], was CPU-optimized and do not fully leverage GPU capabilities.

## 4. Results

In this section, we first verify that the current implementations of all projectors and backprojectors are mutually adjoint operators. Next, we analyze the accuracy of the TT projector and various variants of the CVP projector, comparing them to a densely sampled ray-casting projector based on Siddon's algorithm. We then assess the impact of elevation correction on the accuracy of the CVP projector. Finally, we compare execution times of the projectors on multiple NVIDIA GPUs. Evaluations were performed with Git commit `788ac96` of [25].

### 4.1. Adjoint product test

The implementation in [25], see Table 1, includes for each projector also corresponding backprojector. Given a correctly implemented projection operator, constructing its corresponding backprojection operator is straightforward, however formal verification of their adjointness directly from the code is practically infeasible. Therefore, we employ the following test based on random data, which provides a practical verification of adjointness. Specifically, we use the dot product test using randomly generated volume data $\mathbf{v}$ and a random right-hand side $\mathbf{p}$. For an adjoint operator pair $\mathbf{A}$ and $\mathbf{A}^\top$, the following condition must hold

$$\mathbf{p} \cdot (\mathbf{A}\mathbf{v}) = \mathbf{v} \cdot (\mathbf{A}^\top \mathbf{p}). \quad (41)$$

The adjoint test was performed for each projector and backprojector pair in Table 1. The computed ratio between both sides of Eq. (41) was consistently $1 + \epsilon$, with $|\epsilon| < 10^{-5}$. Since our cutting voxel projector and all other implementations pass this test for random data, we are confident that they correctly form adjoint operator pairs. We cover this behavior in the unit tests of the software.

### 4.2. Comparison of the accuracy

Although projectors can be expressed as linear operators (1), the individual elements of the projector matrix behave highly nonlinearly with respect to voxel position, pixel position, and projection angle. This nonlinearity arises from two independent discretization processes: the subdivision of the reference volume into voxels and the sampling of the detector plane into discrete pixels. The boundaries between these elements introduce additional complexity, making the accurate construction of projectors challenging.

Assessing the accuracy of a projector requires establishing a reliable ground truth. A common approach in the field is to use a densely sampled ray-casting method as a high-fidelity reference. In our case, we construct the raycaster using the Siddon algorithm (see Section 2.5). However, before adopting it as a benchmark, we must ensure that its error remains at least an order of magnitude lower than that of the evaluated projectors. To achieve this precision, we use `Siddon512` as the ground truth, casting 512 × 512 rays per pixel.

We evaluate the projection error of the studied projectors across three different scanning setups, referred to as A, B, and C, which are described in detail in Section 4.2.1. Comparing the studied projectors to a single ray-caster sampling density is not straightforward, as their accuracy varies significantly with elevation angle, see Fig. 4. When the elevation angle is small, very densely sampled ray-casting reference `Siddon128` is needed to get superior accuracy. However, at higher elevation angles, the studied projectors resemble ray-casting projector with coarser sampling, such as `Siddon8`. Since `Siddon128` provides higher accuracy than the studied projectors in all configurations, we choose a ground truth projector `Siddon512` with an accuracy even two orders of magnitude higher. This ensures that the reference remains significantly more precise than the evaluated projectors in all scenarios.

#### 4.2.1. Experimental setup for accuracy evaluation

We evaluate the accuracy of the projectors under three distinct geometrical configurations, labeled as setups A, B, and C. These setups correspond to the top, middle, and bottom graphs in Figs. 4 and 5, respectively. The relative error metric, plotted on the $y$-axis of these graphs, is defined as

$$e_{PRJ} = \frac{\|\mathbf{P}^{PRJ} - \mathbf{P}^{S}\|}{\|\mathbf{P}^{S}\|}, \tag{42}$$

where $\mathbf{P}^{PRJ}$ and $\mathbf{P}^{S}$ represent the projections computed by the evaluated projector and the reference Siddon raycaster, respectively.

**Setup A (Top Graph):** In this configuration, we use a circular scan trajectory with a cone-beam computed tomography (CBCT) setup. The source-to-isocenter distance is $749\,\text{mm}$, and the source-to-detector distance is $1198\,\text{mm}$. The detector matrix consists of 616 × 480 pixels, $0.154\,\text{mm}\times 0.154\,\text{mm}$ each. We analyze a voxel of size $1\,\text{mm}\times 1\,\text{mm}\times 5\,\text{mm}$ positioned at the center of rotation. This setup corresponds to zero elevation angle, typical of conventional CT systems.

**Setup B (Middle Graph):** This setup also employs a circular CBCT trajectory with the same detector and source geometry as Setup A. However, we analyze a smaller voxel of size $1\,\text{mm}\times 1\,\text{mm}\times 1\,\text{mm}$, shifted toward the detector's edge while remaining visible from all views. The voxel center is located at $20\,\text{mm} \times 20\,\text{mm} \times 20\,\text{mm}$. This configuration tests the accuracy of the projector under more challenging conditions, while still maintaining a moderate cone angle of $\sim 2°$.

**Setup C (Bottom Graph):** To evaluate projector performance at high elevation angles, we adopt the setup described in [34]. This configuration uses a circular CT trajectory with a flat-panel detector. The source-to-detector distance is $949\,\text{mm}$, and the source-to-isocenter distance is $541\,\text{mm}$, with 360 uniformly spaced projection views. The detector matrix consists of 768 × 768 detector cells, each measuring $1.0\,\text{mm} \times 1.0\,\text{mm}$. We analyze a voxel positioned at $100\,\text{mm} \times 150\,\text{mm} \times -100\,\text{mm}$, subject to elevation angles $\sim 20°$. This extreme scenario, which exceeds typical CT system conditions, is included to rigorously test the limits of our method.

In all setups, the relative error $e_{PRJ}$ is computed as a function of the projection angle and plotted on either a logarithmic scale (Fig. 4) or a linear scale (Fig. 5) to highlight the accuracy differences between the evaluated projectors and the reference Siddon raycaster. In Fig. 4, the logarithmic scale is used to compare our projectors against a broad range of Siddon raycasters with varying sampling densities, from `Siddon1` to `Siddon256`. In Fig. 5, we investigate more subtle differences between the relaxed and standard versions of the cutting voxel projector, and thus a linear scale is employed.

### 4.3. Comparison of the speed

In this section, we evaluate the computational performance of the projectors on RTX A6000, V100, and A100 NVIDIA GPUs, available on the computational infrastructure we have access to. To compare the speed of the projectors, we created two benchmark problems based on setups from the literature, where the authors reported the speed of the TT projector. While we provide the times reported in these papers for reference, it is important to note that they were obtained in the past on different hardware and under different settings. Our benchmarks aim to provide a fair and consistent comparison of projector and backprojector speeds.

To measure the speed of the projectors and backprojectors, we generated random projection data from a uniform distribution . These data were used as input for 40 iterations of the CGLS reconstruction algorithm. The reported times represent the average projector and backprojector times within the reconstruction, excluding I/O operations such as loading data into GPU memory and reading results back. These one-time operations typically take on the order of seconds, depending on the storage device used.

#### 4.3.1. Experimental setup for speed evaluation

In Tables 2 and 3, we compare the speeds of projectors and backprojectors in two distinct benchmark setups, as described in detail below. The evaluated methods include the cutting voxel projector (CVP), its relaxed version with local memory optimizations (CVP relaxed), the cutting voxel projector without elevation correction (CVP NO EC), its relaxed version (CVP NO EC relaxed), the TT projector, and the `Siddon8` raycaster. These benchmarks provide a comprehensive comparison of computational performance across different hardware configurations and projector implementations.

**Benchmark 1:** The first benchmark is derived from the setup in [34], which uses a circular CT trajectory with a flat detector. The source-to-detector distance is $949\,\text{mm}$, and the source-to-isocenter distance is $541\,\text{mm}$, with 720 equally spaced projection views. The detector matrix consists of 512 × 512 with pixel sizes of $1.0\,\text{mm}\times 1.0\,\text{mm}$. The reconstruction volume has dimensions 512 × 512 × 128 with the voxel sizes of $0.5\,\text{mm} \times 0.5\,\text{mm} \times 0.5\,\text{mm}$.

This results in a tomographic reconstruction problem (1) with $512 \cdot 512 \cdot 128 = 33.5M$ elements of the volume vector $\mathbf{v}$ and $512 \cdot 512 \cdot 720 = 188.7M$ elements of the projection vector $\mathbf{p}$. Long et al. [34, p. 1846, Table II] reported the best TT projector time as $91\,\text{s}$, the best backprojector time as $93\,\text{s}$. Our results for this benchmark are summarized in Table 2.

**Benchmark 2:** The second benchmark is based on the setup in [35], which also uses a circular CT trajectory with a flat detector. The source-to-detector distance is $1000\,\text{mm}$, and the source-to-isocenter distance is $750\,\text{mm}$, with 100 equally spaced views over a $198°$ arc. The detector matrix has dimensions 1280 × 960 with pixels of size $0.25\,\text{mm}\times 0.25\,\text{mm}$. The reconstruction volume has dimensions 256 × 256 × 256 with the voxel sizes of $0.5\,\text{mm}\times 0.5\,\text{mm}\times 0.5\,\text{mm}$.

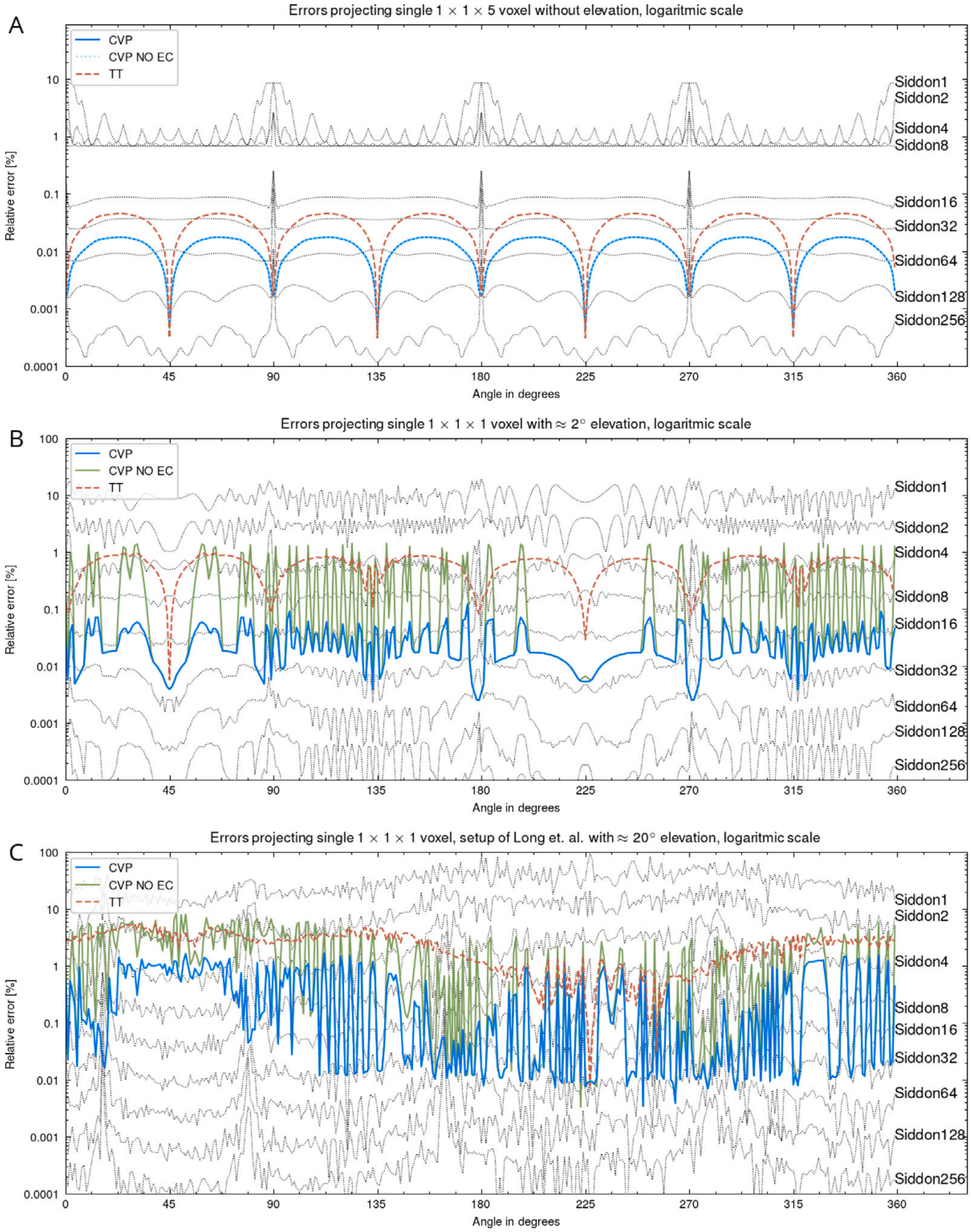

**Fig. 4.** Comparison of projection errors for CVP with and without elevation correction and TT projector against Siddon ray-casting projectors at different sampling densities across three setups described in Section 4.2.1, with relative error on a logarithmic scale.

**Table 2**
Benchmark 1 of projector/backprojector run times, as described in Section 4.3.1.

| | RTX A6000 | | V100 | | A100 | |
|---|---|---|---|---|---|---|
| | P | BP | P | BP | P | BP |
| CVP | 164.3 s | 76.9 s | 33.3 s | 9.8 s | 30.6 s | 5.9 s |
| CVP relaxed | 11.5 s | 4.0 s | 11.0 s | 2.9 s | 8.1 s | 1.9 s |
| CVP NO EC | 90.0 s | 45.3 s | 23.1 s | 6.1 s | 25.6 s | 3.7 s |
| CVP NO EC relaxed | 10.1 s | 3.8 s | 9.6 s | 2.3 s | 7.3 s | 1.5 s |
| TT | 288.5 s | 84.6 s | 66.2 s | 11.2 s | 104.7 s | 6.7 s |
| Siddon8 | 427.8 s | 616.3 s | 60.5 s | 637.5 s | 41.0 s | 142.3 s |

**Table 3**
Benchmark 2 of projector/backprojector run times, as described in Section 4.3.1.

| | RTX A6000 | | V100 | | A100 | |
|---|---|---|---|---|---|---|
| | P | BP | P | BP | P | BP |
| CVP | 121.6 s | 56.9 s | 15.5 s | 6.8 s | 15.1 s | 4.1 s |
| CVP relaxed | 9.1 s | 2.8 s | 7.0 s | 2.0 s | 6.1 s | 1.3 s |
| CVP NO EC | 68.5 s | 34.0 s | 11.3 s | 4.1 s | 11.8 s | 2.6 s |
| CVP NO EC relaxed | 7.7 s | 2.7 s | 5.9 s | 1.3 s | 5.2 s | 0.9 s |
| TT | 130.0 s | 52.2 s | 16.4 s | 6.4 s | 18.4 s | 4.1 s |
| Siddon8 | 905.5 s | 1403.3 s | 94.3 s | 1183.4 s | 68.2 s | 291.5 s |

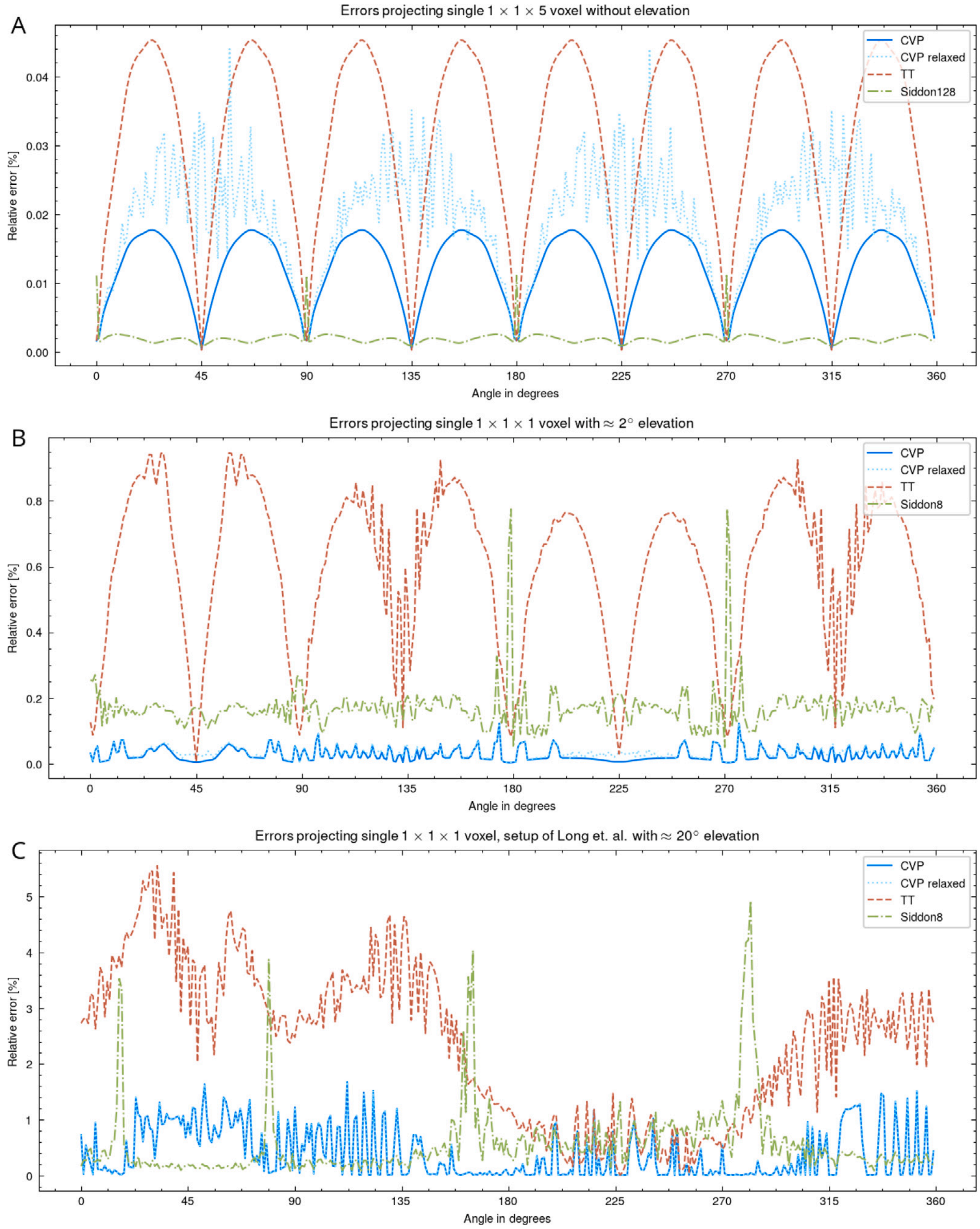

**Fig. 5.** Comparison of projection errors for CVP with elevation correction, its relaxed version, TT projector, and a selected Siddon projector across three different setups (A, B, and C) described in Section 4.2.1. The relative error is shown on a linear scale.

This results in a tomographic reconstruction problem (1) with $256\cdot 256\cdot 256 = 16.8M$ elements of the volume vector $\mathbf{v}$ and $1280\cdot 960\cdot 100 = 122.9M$ elements of the projection vector $\mathbf{p}$. Pfeiffer et al. [35, Table 1] reported the time of the TT projector as $46.3\,\text{s}$, backprojector time was not disclosed. Our results for this benchmark are summarized in Table 3.

## 5. Discussion

Our contribution focuses on the design and implementation of a novel voxel-driven projector for cone beam computed tomography (CT) that computes pixel contributions using the volumes of voxel cuts. It results in a near-exact projector derived directly from the physical principles of ray attenuation and voxel–pixel interactions. The mathematical formulation is detailed in Section 2, where we demonstrate how this approach overcomes the limitations of phenomenological models, offering enhanced accuracy, especially for complex cone-beam geometries.

We also present a GPU-optimized implementation of the projector in Section 3, including a simplified algorithm for calculating cuts as detailed in Section 3.1. Our implementation features efficient local memory write synchronization across multiple computational units, which significantly accelerates computation while maintaining high accuracy. The effectiveness of these algorithmic improvements is further validated in Section 4, where we compare the accuracy and speed of various projector variants.

### 5.1. Elevation correction

In Fig. 4 it can be seen that the version of the projector without elevation correction exhibits peaks in error that occasionally exceed those of the TT projector, this phenomenon is not observed at zero elevation angles. Upon investigating these discrepancies, we determined that the issue does not stem from the foundational formulas of the projector (35) and (37), but rather from the current method used to compute volume cuts. Specifically, voxel cuts in the detector's orthogonal direction $\chi_2$ are computed as orthogonal projections, whereas a more natural approach would employ skew projections that incorporate the appropriate skew angle corresponding to the elevation.

To address this issue, we implemented so called elevation correction see Section 3.1.1. Due to intra-voxel compensatory effects, we apply the elevation correction selectively, only at those pixels onto which the top and bottom of the voxel cuts project, leaving intermediate pixels unchanged. This targeted approach maintains low computational costs while ensuring that the projector's accuracy remains consistently below that of the TT projector, as evidenced by Fig. 4.

In practical computations, both the TT projector and the CVP (with and without elevation correction) yield very good results, with differences often being negligible due to the similarity of neighboring voxel values. While the elevation correction provides enhanced accuracy, especially in high-precision scenarios, its additional computational cost is modest. Speed comparisons for the relaxed CVP versions in Tables 2 and 3 show that applying the elevation correction increases computational cost by approximately 10%–25% compared to the non-corrected implementations. This quantification helps to clarify the trade-offs involved and demonstrates that the correction remains applicable across a broader range of applications despite the slight increase in computational overhead.

### 5.2. Speed and GPU optimization

Tables 2 and 3 provide a comprehensive comparison of the CVP and its relaxed variant, both with and without elevation correction. For reference, these tables also include execution times for the TT projector and the relatively densely sampled raycaster `Siddon8`. The speed improvements observed in the relaxed version are attributable, in part, to the switch from `float64` to `float32` precision and the use of `cl-fast-relaxed-math`, as detailed in Section 3.2. Additionally, further speed gains in the projector implementation arise from optimized local memory operations and synchronization strategies, as described in Section 3.3.

In terms of computational overhead, the cost of incorporating elevation correction is on the order of 10%–25%. In contrast, the relaxed implementations are between 2.5 and 15 times faster than the non-relaxed implementations, with the exact improvement depending on the underlying GPU architecture. Furthermore, as demonstrated in Fig. 5, the errors introduced by the relaxed version are negligible, especially at higher elevation angles. Even at zero elevation angle, where the differences are visible in Fig. 5, the absolute error difference remains minimal (around 0.02%).

### 5.3. Conclusion and future work

Since its initial implementation by V. Kulvait in 2019, the cutting voxel projector (CVP) has been successfully applied to numerous reconstruction problems, including advanced CBCT perfusion data processing, as documented in previous works [17,20]. More recently, the underlying concept has been extended to parallel-ray geometries in synchrotron $\mu$-CT setups, demonstrating the method's versatility and broad applicability in tomographic imaging.

Despite optimizations for concurrent pixel value updates in local GPU memory, the forward projector remains slower than the backprojector. For speed-sensitive applications, it may be preferable to use a fast, low-accuracy ray-tracing method for the forward projection, while retaining a more precise backprojector to avoid amplifying errors in iterative reconstructions. This strategy is employed in frameworks like ASTRA [22], which combine a forward raytracer with a backward footprint projector for improved speed, albeit at the expense of losing self-adjointness of the operator.

Looking forward, our derivation and implementation of the CVP provide a strong foundation for future research. One promising direction is the extension of the CVP to non-circular scanning trajectories. Although our current implementation in Section 3 assumes alignment between the voxel $x_3$ axis and the detector $\chi_1$ axis, the theoretical derivation in Section 2 is more general. Combining these results with algorithms for computing intersections of cubes with half-planes, e.g. clipping algorithms, it should be possible to develop projectors adaptable to arbitrary scanning geometries, overcoming a key limitation of footprint projectors, which inherently depend on a fixed geometric alignment.

One notable feature of the CVP is its natural compatibility with hierarchical reconstruction using non-uniform voxel grids, as it operates independently of a voxel size. This flexibility allows for the development of advanced computational techniques analogous to algebraic multigrid (AMG) methods in partial differential equations. By enabling local voxel refinement, it becomes possible to achieve higher resolution in regions of interest while maintaining a coarser resolution elsewhere, improving reconstruction efficiency without sacrificing detail in critical areas.

In summary, the cutting voxel projector not only delivers improved accuracy and runtime performance but also establishes a versatile platform for future innovations in adaptive tomographic reconstruction. The work presented here lays the groundwork for further explorations into hierarchical solvers and more flexible scanning geometries, promising to enhance the scalability and efficiency of algebraic reconstruction techniques in increasingly complex imaging scenarios.

## CRediT authorship contribution statement

**Vojtěch Kulvait:** Writing – review & editing, Writing – original draft, Software, Methodology, Formal analysis, Conceptualization. **Julian Moosmann:** Supervision, Resources, Funding acquisition. **Georg Rose:** Supervision, Resources, Funding acquisition.

## Declaration of Generative AI and AI-assisted technologies in the writing process

During the review of this work authors used ChatGPT in order to improve readability and the flow of the text. After using this tool/service, authors reviewed and edited the content as needed and takes full responsibility for the content of the publication.

## Declaration of competing interest

The authors declare that they have no known competing financial interests or personal relationships that could have appeared to influence the work reported in this paper.

## Acknowledgments

V. Kulvait (VK) designed and implemented the cutting voxel projector for the CBCT operator during his postdoctoral position at OVGU. Later, while serving as a scientist at Hereon, he developed a simplified version for parallel beam geometry tailored for synchrotron $\mu$-CT applications. The OpenCL kernels are now included in the open-source package for tomographic reconstruction, KCT_cbct [25]. The implementations are publicly available as open-source software under the GNU GPL 3 license at https://github.com/kulvait/KCT_cbct.

## Data availability

Data will be made available on request.